\documentclass[aps,prb,psfig,amstex,showpacs,onecolumn]{revtex4}
\usepackage{amssymb}
\usepackage{amsmath}
\usepackage{graphicx}

\begin{document}
\title{Comment on
"Single-parameter quantum charge and spin pumping in armchair graphene nanoribbons" (arXiv:1206.3435v1, by Zhou  and Wu) }

\author{Zhen-Gang Zhu and Jamal Berakdar}
\affiliation{
Institut f\"{u}r Physik, Martin-Luther-Universit\"{a}t Halle-Wittenberg,  06099 Halle (Saale), Germany}
%$^2$Department of Physics and Electronic Science, Changsha University of Science and Technology, Changsha,410076, China}

\begin{abstract}
In their recent submission 	arXiv:1206.3435v1   Zhou  and Wu  addressed the charge and spin pumping in an armchair graphene nanoribbon connected to
magnetic and/or nonmagnetic leads at zero bias. They used thereby a pumping current formula based on time-reversal symmetry.
In their appendix they made, based on their formulas,  a number of  claims concerning our recent works \cite{ding} on the laser modification of the
spin-dependent  current in an extended graphene monolayer contacted to two magnetic leads under a finite bias.
Below we show in detail that the physical system considered by us is conceptually different from that of Zhou and Wu (arXiv:1206.3435v1) and cannot be treated with their theory for fundamental reasons
(e.g., the time-reversal symmetry is inherently broken in our case due to the presence of the magnetic leads and the dc current generated by the
finite bias). Hence, the statements of
Zhou  and Wu  based on a comparison of their results with ours are scientifically groundless.
\end{abstract}

%\pacs{65.80.Ck, 72.80.Vp, 73.50.Lw }

\maketitle

In the submission 	arXiv:1206.3435v1   Zhou  and Wu   presented their views on the spin pumping in an armchair graphene nanoribbon under an  ac gate voltage while  the nano ribbon is being connected to nonmagnetic/ferromagnetic leads.  Such an approach cannot make any profound statements on the
laser-induced modulations of the dc current in an extended graphene sheet contacted to two biased magnetic leads, a case which we studied in
\cite{ding,ding1}. Hence, their claims and statements on our calculated current as a function of the applied
bias lack a scientific basis.

Here we only focus on the difference between our case and the case studied by  Zhou  and Wu.
Their starting point is eq.(17) in arXiv:1206.3435v1 for the pumping current in the spin channel $\sigma$
\begin{equation}
I^{\sigma}_{\text{pump}}=\frac{e}{h}\sum_{n}\int_{-\infty}^{E_{F}}d\varepsilon\left[T_{LR\sigma}^{n}(\varepsilon)-T_{RL\sigma}^{n}(\varepsilon)\right].
\label{current0}
\end{equation}
 $T_s$ are the transmission probabilities in standard notations, $E_F$ is the Fermi  energy, and $e$ is the electron charge.
 As stated in their paper, they use a \emph{time-reversal symmetry} (TRS) to write $T_{LR\sigma}^{n}(\varepsilon)=
 T_{LR\sigma}^{-n}(\varepsilon+n\Omega)$ where $\Omega$ is the frequency of the ac field applied to the  graphene nano ribbon.
 With this they proclaim  (Eq. (18)  arXiv:1206.3435v1) that
\begin{equation}
I^{\sigma}_{\text{pump}}=\frac{e}{h}\sum_{n>0}\int_{E_{F}-n\Omega}^{E_{F}}d\varepsilon\left[T_{LR\sigma}^{n}(\varepsilon)-T_{RL\sigma}^{n}(\varepsilon)\right].
\label{current}
\end{equation}
is an exact formula valid without the need  to introduce an energy cutoff because the upper limit is $E_{F}$  and the lower limit is $E_{F}-n\Omega$.
On the  basis of this formula
they show in their Fig. 5(a) two sets of curves: one set is generated by Eq. (\ref{current}) and one set is generated by introducing a theta function $\Theta(E_{\text{cut}}-|\varepsilon+N\Omega|))$ in the Green function that enters, e.g. $T_s$ (here $E_{\text{cut}}$ is an ad hoc energy cutoff).
In Fig. 5(b), they plot the \textit{pumping} current vs. $V_{\text{ac}}$ by using the two ways. By the latter (cutoff) way, they notice that the current at first increases and then slowly decreases. They claimed that these curves generated by the latter way coincide our results  and further claim
 that the current decrease they observe in their calculations
  is just a consequence of an introduced cutoff. In fact, they compare their own calculations generated  by their two different equations.
  A comparison with our results has absolutely no
scientific ground. Their approach does not apply to our system (as a matter of fact nor to any system with a broken time-reversal symmetry, e.g.
when magnetic leads or dc currents are involved) and in particular, their scheme does not reproduce our formulas.

%We show here their conclusions about our work are completely wrong. Firstly, one should really repeat the target results which one wants to criticize. Only in this way can the results be thought of whether being suffered a problem. However, the curves in Fig. 5 are not our results. These figures are only the consequences of their formulas and their treatments. The reasons are given in the following.
For clarity we note the following:

\begin{enumerate}
\item
In our work, we did not study  pumping effects. There is always a dc bias between the two magnetic leads
 in addition to  the ac field acting on the graphene sheet. This dc bias gives rise to a background current. We investigated the effect of the ac field on the background current. While, Zhou and Wu's work is about the charge pumping and spin pumping. There is no such a dc bias through the system. Their formulas are not applicable to our system. This is why they call the calculated current in Fig. 5  pumping current. Therefore, their demonstration and conclusions have nothing to do with ours.

\item
They claim their formula, Eq. (18) for the pumping current has the benefit that no cutoff is needed. They derive this formula by using the crucial relation
\begin{equation}
T_{LR\sigma}^{n}(\varepsilon)=T_{RL\sigma}^{-n}(\varepsilon+n\Omega).
\label{trs}
\end{equation}
They claimed that this is due to the time-reversal symmetry (TRS). We should point out that in our study we considered a system with ferromagnetic (FM) leads where spin splitting is present. In the presence of these FM leads, there is no TRS. One can define a time reversal operator and try to calculate the commutator with the Hamilton operator including the FM leads. The finding is that these operators do not commute. Therefore, there is no TRS. The current formula, i.e. Eq. (18) in their paper (i.e. Eq. (\ref{current}) here), needs to be justified fundamentally.
 Their demonstrations for the spin dependent pumping  currents should thus be taken with care, but anyway they do not compare
 with our results because we are dealing with a different physical problem.
\item
The TRS consideration, i.e. the Eq. (\ref{trs}), is not straightforward. Usually, the TRS is verified by constructing a time reversal operator and considering its commutation properties with the Hamiltonian. The demonstration of TRS can not be done based on a
 consideration of a calculated statistical physical quantity. For example,  when a dc bias is present (as in our study), the system is in a nonequilibrium situation and a charge current is flowing through the system (the leads are expected to act as particle baths).  TRS does not exist in this case even for systems with nonmagnetic leads.
 %
 %When the quantum statistics enters into the calculation, can we still think there is a TRS? Can we demonstrate the TRS by using a physical quantity, for example the transmission probability? When a dc bias is present (in our study), the system is in nonequilibrium situation and current is flowing through the system. The TRS does not exist in this case even for the systems with nonmagnetic leads.
%
\item
Zhou and Wu studied the time-averaged current through a lead/graphene nanoribbon/lead system in which an ac field is applied to the graphene region. Under such an ac field, the graphene part  is actually in a nonequilibrium state. To calculate the current, we have to start from the charge conservation law \cite{haug}
\begin{equation}
e\frac{dN_{\text{G}}}{dt}=J_{\text{R}}(t)+J_{\text{L}}(t),
\label{chargeconservation}
\end{equation}
where $N_{\text{G}}$ is the occupation of graphene, $J_{\text{L}}(t)$ and $J_{\text{R}}(t)$ are the current flowing through the left and the right leads. The left hand side of the equal sign is just the displacement current reflecting the charge accumulation in the central region and related to the leads currents. We can use the standard definition of the time-averaged current \cite{haug}
\begin{equation}
\langle F(t)\rangle=\lim_{T\rightarrow\infty}\frac{1}{T}\int_{-T/2}^{T/2}dtF(t).
\label{timeaverage}
\end{equation}
The displacement current tends to zero after this time average. Only in this case, will the charges flowing from the left lead, flow into the right lead completely. There is no charge accumulation in the central region in this case. However, Zhou and Wu used another definition, namely
\begin{equation}
\bar{I}=\frac{1}{T_{0}}\int_{0}^{T_{0}}I,
\label{time2}
\end{equation}
where $T_{0}=2\pi/\Omega$. This is an average in \textit{one period} of the ac field. In this time period, the displacement current is in general finite  and there is a charge accumulation in the central region. However,  they ignore this term in their Eq. (6). A serious problem is that the charge conservation is violated in the formula of the current. Therefore, we conclude that the equations of the current
and the calculations based upon these equations need to be reexamined

\end{enumerate}

In summary, the remarks and statements by Zhou and Wu's  based on a comparison of their calculations in arXiv:1206.3435v1
with our works  \cite{ding,ding1} have no scientific basis.

%\newpage

%\textbf{Acknowledgements:}
%Z. G. Z. and J. B. are supported by DFG. K.H.D. is supported by  DAAD (Germany).

%\newpage

%\appendix
%

\end{document}